\documentclass[traditabstract, letter, longauth]{aa}

\usepackage{txfonts}
\usepackage{graphicx}
\usepackage{natbib}
\usepackage{amssymb}

\begin{document}

\title{{\it Herschel}-SPIRE observations of the Polaris flare : structure of the diffuse interstellar medium at the
sub-parsec scale\thanks{{\it Herschel} is an ESA space observatory with science instruments provided 
by European-led Principal Investigator consortia and with important participation from NASA.}}

\author{
M.-A. Miville-Desch\^enes\inst{1,14}, P. G. Martin\inst{14}, A. Abergel\inst{1}, J.-P. Bernard\inst{7}, 
F. Boulanger\inst{1}, G. Lagache\inst{1},
 L. D. Anderson\inst{2},
 P. Andr\'e\inst{4},
 H. Arab\inst{1},
 J.-P. Baluteau\inst{2},
 K. Blagrave\inst{14},
 M. Cohen\inst{9},
 M. Compiegne\inst{14},
 P. Cox\inst{10},
 E. Dartois\inst{1},
 G. Davis\inst{11},
 R. Emery\inst{17},
 T. Fulton\inst{20},
 C. Gry\inst{2},
 E. Habart\inst{1},
 M. Huang\inst{11},
 C. Joblin\inst{7},
 S. C. Jones\inst{16},
 J. Kirk\inst{3},
 T. Lim\inst{17},
 S. Madden\inst{4},
 G. Makiwa\inst{16},
 A. Menshchikov\inst{4},
 S. Molinari\inst{15},
 H. Moseley$^{19}$,
 F. Motte\inst{4},
 D. A. Naylor\inst{16},
 K. Okumura\inst{4},
 D. Pinheiro Gocalvez\inst{14},
 E. Polehampton$^{16,17}$,
 J. A. Rod\'on\inst{2},
 D. Russeil\inst{2},
 P. Saraceno\inst{15},
 N. Schneider\inst{4},
 S. Sidher\inst{17},
 L. Spencer\inst{16},
 B. Swinyard\inst{17},
 D. Ward-Thompson\inst{3},
 G. J. White\inst{17,22},
 A. Zavagno\inst{2}
  }

   \institute{
 Institut dÕAstrophysique Spatiale, CNRS/Universit\'e Paris-Sud\,11, 91405 Orsay, France  \and
 Laboratoire d'Astrophysique de Marseille (UMR\,6110 CNRS \& Universit\'e de Provence), 38 rue F.  
 Joliot-Curie, 13388 Marseille Cedex 13, France \and
 Department of Physics and Astronomy, Cardiff University, Cardiff, UK \and
 CEA, Saclay, France \and
 Department of Physics and Astronomy, University College London, London  UK \and
 IPAC, California Institute for Technology, Pasadena, USA \and
Universit\'e de Toulouse ; UPS ; CESR ; CNRS ; UMR5187 ; 9 avenue du colonel Roche,
F-31028 Toulouse cedex 4, France \and
 Observatoire de Bordeaux, France \and
 Radio Astronomy Laboratory, University of California, Berkeley, USA \and
 IRAM, Grenoble, France \and
 NOAC, China \and
National Research Council of Canada, Herzberg Institute of Astrophysics, Victoria, Canada \and
 Department of Physics and Astronomy, University of British Columbia, Vancouver, Canada \and
 CITA, Toronto, Canada \and
 CNR - Istituto di Fisica dello Spazio Interplanetario, Roma, Italy \and
 Institute for Space Imaging Science, University of Lethbridge, Lethbridge, Canada \and
 Space Science Department, Rutherford Appleton Laboratory, Chilton, UK \and
 Centre for Astrophysics and Planetary Science, School of Physical Sciences, University of Kent, Kent, UK \and
  NASA - Goddard SFC, USA \and
 Blue Sky Spectrosocpy Inc, Lethbridge, Canada \and
Laboratoire des Signaux et Systmes, SUPELEC , Plateau de Moulon, 91192 Gif-sur-Yvette Cedex, France \and
Department of Physics \& Astronomy, The Open University, Milton Keynes MK7 6AA, UK}

 \offprints{Marc-Antoine Miville-Desch\^enes}
 \mail{mamd@ias.u-psud.fr}
 \date{\today}

\titlerunning{{\it Herschel}-SPIRE observations of the Polaris flare}
\authorrunning{Miville-Desch\^enes, M.-A. et al.}

\abstract{We present a power spectrum analysis of the {\it Herschel}-SPIRE observations of the Polaris flare, 
a high Galactic latitude cirrus cloud midway between the diffuse and molecular phases. 
The SPIRE images of the Polaris flare reveal for the first time the structure of the diffuse interstellar medium 
down to 0.01 parsec over a 10 square degrees region. 
These exceptional observations highlight the highly filamentary and clumpy structure of the interstellar medium 
even in diffuse regions of the map. 
The power spectrum analysis shows that the structure of the interstellar medium is well described by a single power law 
with an exponent of $-2.7\pm0.1$ at all scales from 30'' to 8$^\circ$.
That the power spectrum slope of the dust emission is constant down to the SPIRE angular resolution is an indication that
the inertial range of turbulence extends down to the 0.01~pc scale.
The power spectrum analysis also allows the identification of a Poissonian component 
at sub-arcminute scales in agreement with predictions of the cosmic infrared background level at SPIRE wavelengths.
Finally, the comparison of the SPIRE and IRAS~100$\mu$m data of the Polaris flare clearly
assesses the capability of SPIRE in maping diffuse emission over large areas.
}


\maketitle

\section{Introduction}

Understanding the physical mechanisms involved in the formation of stars is still one of the major aspirations 
of modern astrophysics. One of the many aspects of this problem is related to the organisation of matter
in the diffuse interstellar medium (ISM) as its structure provides the initial conditions for the formation of
molecular clouds.
The physics that impacts on the structure of the ISM is complex : 
it involves compressible and magnetic turbulent motions in a thermally unstable flow. 
Even though turbulence organises the flow up to scales of tens of parsec {\cite[]{elmegreen2004} }, 
important non-linear physical processes take place at the sub-parsec scales. 
Being able to study the kinematics of the gas at such scale is essential 
but the density structure is also a key tracer of the physics involved. 

Dust emission from uniformely heated regions of the local interstellar medium 
{(i.e. exempt of local production of UV photons by stars)}
allows us to directly trace the density structure of the ISM. 
With its unprecedent angular resolution, the ESA {\it Herschel} Space Observatory
\cite[]{pilbratt2010} allows us for the first time to study the structure of the ISM 
at the 0.01~pc scale over several square degrees.
We present a first analysis of the 250, 350 and 500~$\mu$m emission maps of 
the Polaris flare obtained with the SPIRE instrument \cite[]{griffin2010}. 
These observations are part of the 
``Evolution of Interstellar dust''  \cite[]{abergel2010} and ``Gould Belt survey'' \cite[]{andre2010} 
SPIRE key projects.

The Polaris flare is a high Galactic latitude cirrus cloud, located at
a distance $\leq 150$~pc \cite[]{falgarone1998}. It has significant CO emission
 \cite[]{heithausen1990,meyerdierks1996,falgarone1998} 
and colder dust grains than typical diffuse clouds \cite[]{lagache1998,bernard1999}.
As the Polaris flare does not show any sign of star-formation activity 
{(only pre-stellar cores are detected with SPIRE - \cite{andre2010}}, 
it is the archetype of the initial phases of molecular cloud formation. 
{We present a power spectrum analysis of the dust continuum emission 
from big grains using SPIRE observations. These data allow us to study 
homogeneously the structure of the diffuse ISM from 0.01 to 8~pc.}

\begin{figure*}
\begin{center}
\includegraphics[width=0.86\linewidth, draft=false, angle=0]{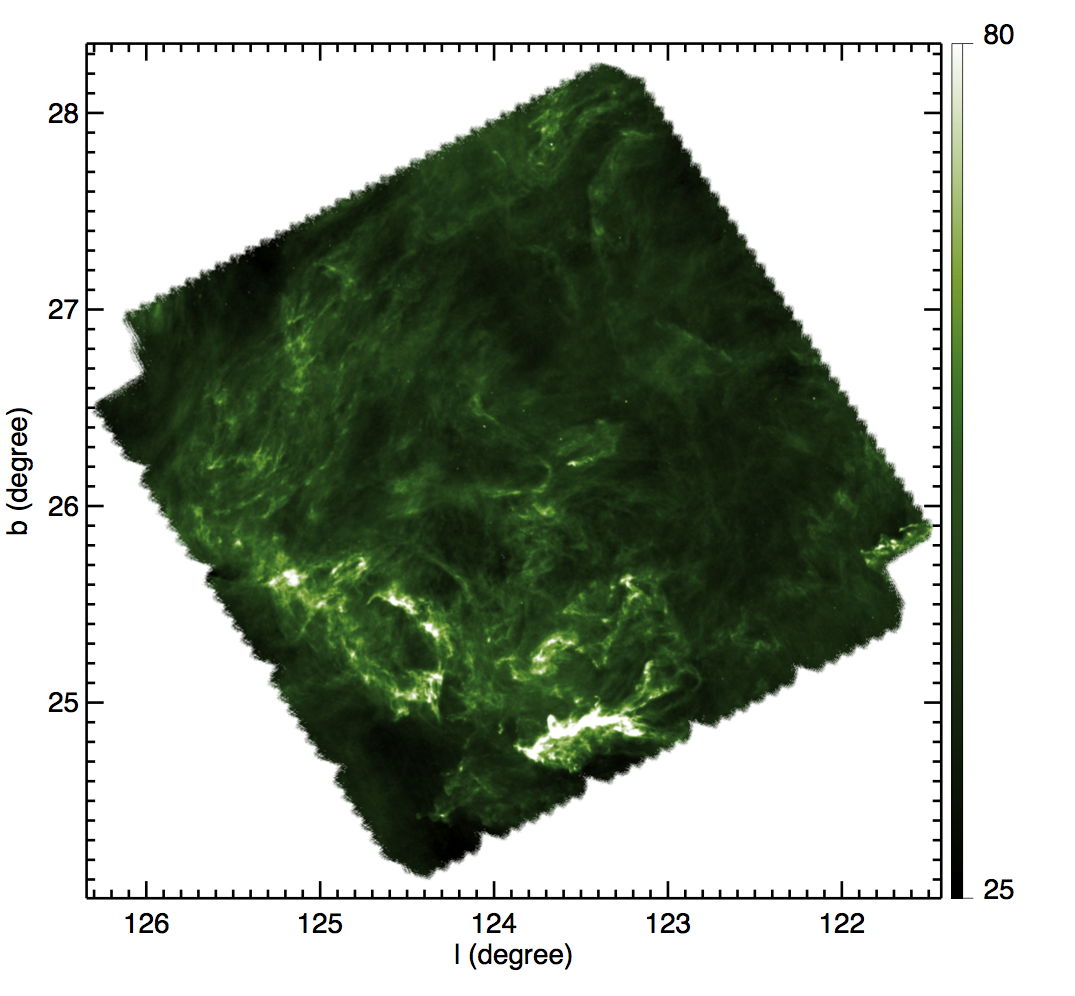}
\caption{\label{fig:250map} SPIRE 250~$\mu$m map of the Polaris flare. 
Units are in MJy~sr$^{-1}$. The zero level was set by correlation with the IRAS/IRIS 100~$\mu$m data. 
}
\end{center}
\end{figure*}

\section{SPIRE observations of the Polaris flare}

The 250~$\mu$m SPIRE map, covering 10 square degrees of the brightest part of the Polaris flare, 
is shown in Fig.~\ref{fig:250map}. The field was observed twice
in nearly perpendical directions to optimize the restoration of diffuse emission.
The SPIRE data were processed with HIPE (version 2.0) applying 
standard corrections for instrumental effects and glitches. The 1/f noise component was removed 
using the "temperature drift correction" module, and naive maps were computed.  

This map reveals for the first
time the structure of the diffuse interstellar medium on scales ranging from 0.01 to 8~pc. Compared to the
previous vision of the diffuse interstellar medium given by lower resolution observations (e.g. IRAS) these
observations reveal a structure with strong contrast at small scales. Numerous small scale clumps
are seen in the map even in the most diffuse regions (see examples in the Appendix). This high-resolution view of the
diffuse ISM also reveals its highly filamentary structure with narrow threads of matter following the larger scale organisation.
These observations bring new insights into the small scale structure of the ISM, and they will certainly help
understand the physical processes dominating the dynamical evolution of matter 
towards the formation of stars. This task is obviously beyond the scope of the present paper.

\section{Comparison with IRAS 100~$\mu$m: check of diffuse emission restoration and dust spectrum}

\label{section:iras}

The estimate of the power spectrum of the interstellar medium emission can only be done with observations 
that restore the power observed at all scales. This implies a great control of instrumental effects that could affect 
the baseline (additive effect) or the gain (multiplicative effect)
of the detectors over the whole period of the observations.
In order to assess the quality of the diffuse emission restoration by SPIRE over such a large field we made a comparison 
over the whole field with the 100~$\mu$m IRAS (IRIS) data \cite[]{miville-deschenes2005a}
which are known to have a good description of the interstellar emission at all scales\footnote{in IRIS the variation with scale of the IRAS 
detectors gain was corrected and the emission at scales larger than 30' was made consistent with DIRBE, 
which was designed for full control on systematic effects.}.
The main limitation of this exercise is the difference in wavelength between SPIRE and IRAS, but
even though local variations of the dust emission spectrum are expected, the fact that
both datasets are dominated by the emission from the big grain population is instructive.

We performed the following linear regression fit :  $S(\lambda) = G\times S(100) + S_0$, where
$S(100)$ is the 100~$\mu$m IRAS/IRIS map from which the average value of the cosmic infrared background at 100~$\mu$m 
(0.7 MJy/sr - \cite{miville-deschenes2007a}) was removed, and $S(\lambda)$ is the SPIRE map at wavelength $\lambda$, 
convolved to the IRAS resolution (4.3 arcmin) and projected onto the native 1.5' grid of IRAS. 
The regression coefficients $G$ and offsets $S_0$ found at each SPIRE wavelength are given in Table~\ref{table:polaris}.
Even though the correlations are good (correlation of 0.85), there is significant variation around the linear fit.
Looking at the difference map (i.e. $S(\lambda) - G\times S(100)$) localized variations are seen, 
which reflect expected modifications of the dust emission spectrum. These small and intermediate scale variations sit
on a fainter large scale structure uncorrelated with the emission and therefore probably 
unrelated to variations of dust properties. 
This residual, which is less than 10\% of the large scale emission fluctuations,
could be attributed to residual imperfections in the data processing. 

\begin{figure}
\includegraphics[width=\linewidth, draft=false, angle=0]{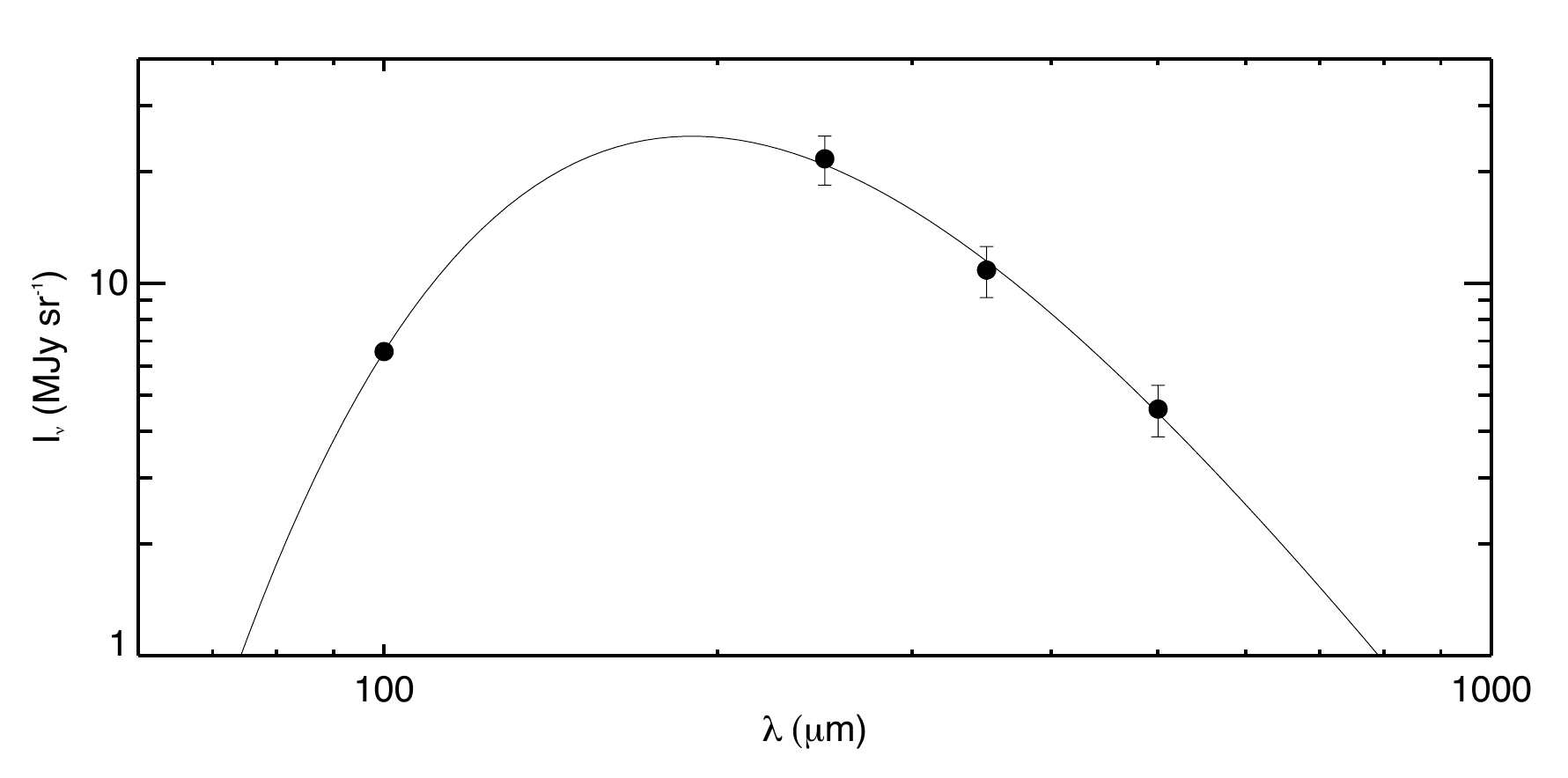}
\caption{\label{fig:avgSED} Average big dust emission spectrum in the Polaris flare as determined by correlation
of the SPIRE and 100~$\mu$m data over the whole field. 
In this plot the correlation coefficients ($G$ in Table~\ref{table:polaris}) were 
scaled to the average 100~$\mu$m brightness in the SPIRE field.
The fit of the big grain emission spectrum gives $T_d=14.5\pm1.6$ and $\beta=2.3\pm0.6$.  
}
\end{figure}

The factors $G$ obtained for the SPIRE-IRAS correlation can be used to estimate the average dust emission 
spectrum in the field. The fit of a grey body\footnote{$I_\nu \propto B_\nu(T_d)\nu^{-\beta}$, where $T_d$ and $\beta$ are the big grain
temperature and emissivity index, and $B_\nu$ is the Planck function.} 
to the IRAS-SPIRE correlation coefficient shown in Fig.~\ref{fig:avgSED} gives 
$T_d=14.5\pm1.6$ and $\beta=2.3\pm0.6$, in agreement with what was measured by 
\cite{bernard1999} using PRONAOS and ISOPHOT data on a $30'\times 6'$ region in the brightest part of the Polaris flare
($T_d=13.0\pm0.8$ and $\beta=2.2\pm0.3$). 
This provides a first order sanity check of the quality of the SPIRE gain calibration. 
In addition, looking at $(S(\lambda)-S_0)/S(100)$, we find no systematic correlation of $G$ with intensity
which agrees with the fact that the SPIRE diffuse emission calibration is not scale-dependent 
at scales larger than the IRAS beam. 

\begin{table*}
\begin{center}
\begin{tabular}{c|c|c|c|c|c|c|c}\hline
$\lambda$ & pixel size & FWHM & $G$ & $S_0$ & noise & $\gamma$ & $P_{0}$ \\
($\mu$m) & (arcsec) & (arcsec) & & (MJy sr$^{-1}$) & (MJy sr$^{-1}$) & & (Jy$^2$ sr$^{-1}$) \\ \hline
250 & 6 & 18.1 & 3.3$\pm$0.5 (0.01) & 20.1$\pm$3.1 (0.1) & 1.26 & $-2.65\pm0.10$ & 5$\pm 2\times 10^3$\\
350 & 10 & 25.2 & 1.7$\pm$0.3 (0.008) & 10.1$\pm$1.6 (0.05) & 0.55 & $-2.69\pm 0.13$ & 4$\pm 2 \times 10^3$\\
500 & 14 & 36.9 & 0.7$\pm$0.1 (0.003) & 4.3$\pm$0.7 (0.02) & 0.34 & $-2.62\pm 0.17$ & 1$\pm 1 \times 10^3$\\ \hline
\end{tabular}
\end{center}
\caption{\label{table:polaris} {\bf SPIRE observations of the Polaris flare. }
Column 4 and 5: gain and offset coefficients of the SPIRE-IRIS~100~$\mu$m correlations. 
The uncertainty on $G$ represents the rms of the ratio 
$S(\lambda)/S(100\mu m)$ once the offset $S_0$ is removed from the SPIRE data. Similarly the uncertainty
on $S_0$ is the rms of $S(\lambda) - G\times S(100\mu m)$. 
These two uncertainties are correlated, but they give more realistic estimates compared 
to the statistical ones obtained with the linear regression fit (given in brackets).
Column 6: noise level estimated directly on the power spectrum for $k>0.75k_{max}$ ($3.75<k<5$~arcmin$^{-1}$ at 250~$\mu$m).
Column 7: power spectrum exponent of the interstellar component.
Column 8: level of the Poissonian component associated with point sources and the CIB fluctuations.
The fit of the power spectrum ($\gamma$ and $P_0$) was done between k=0.025~arcmin$^{-1}$ and about twice the 
FWHM (2.0, 1.2 and 0.86 arcmin$^{-1}$ at 250, 350 and 500~$\mu$m respectively).
}
\end{table*}

\section{Power spectrum analysis}

\begin{figure}
\includegraphics[width=\linewidth, draft=false, angle=0]{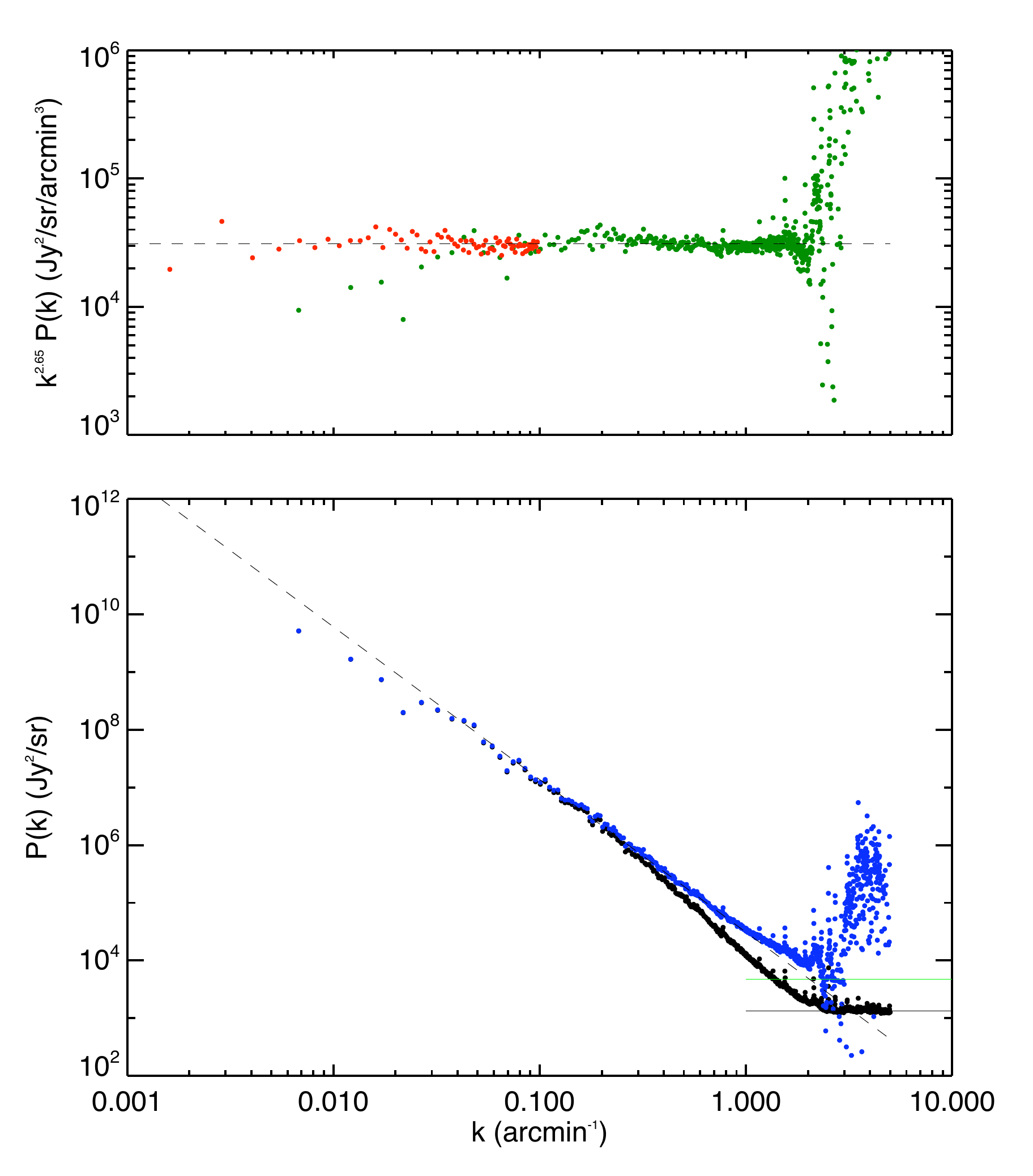}
\caption{\label{fig:powspec250} Power spectrum of the SPIRE 250~$\mu$m map of the Polaris flare. 
{\bf Bottom:} The black dots are
the raw power spectrum (computed with an apodization of 0.97).  The horizontal black line is the white noise level estimate.
Blue dots are the power spectrum noise removed and divided by the psf estimated on Neptune observations. 
The green horizontal line and the dashed line are
the source and interstellar components estimated from the blue dots power spectrum respectively, on scales
$0.025<k< 2$~arcmin$^{-1}$. {\bf Top: } The green dots are the power spectrum of the interstellar component (i.e. corrected for 
noise, psf and sources) multiplied by $k^{2.65}$. 
The red dots are the power spectrum of the IRAS/IRIS 100~$\mu$m emission in a $12^\circ\times12^\circ$ region
centered on the SPIRE field. The dashed line is the same as in the bottom figure. 
}
\end{figure}

The power spectrum of the 250~$\mu$m map of the Polaris flare, converted to $Jy^2/sr$ is shown in Fig.~\ref{fig:powspec250}.
The black dots in the bottom plot show the power spectrum computed on a $2.8^\circ\times3.1^\circ$ area of the map
where all data points are defined and from which bright point sources were removed. 
An apodization factor of 0.97 was applied prior to the Fourier Transform \cite[]{miville-deschenes2002b}.

The power spectrum is typical of infrared emission of high Galactic latitude fields 
with a power-law type spectrum convolved by the instrument transfer function ($\phi$), and a flat noise part ($N$) at high $k$.
The power spectrum is modeled accordingly : 
$P(k) = \phi(k) P_{sky}(k) + N(k).$
The white noise term stands out very clearly in all power spectra. 
Its level is estimated as the average of $P(k)$ for $0.75k_{max}<k<k_{max}$, where $k_{max}$ is the maximum $k$ available
(i.e. twice the pixel size - see Table~\ref{table:polaris}). 
At each SPIRE wavelength this corresponds to scales smaller than the beam size where the noise dominates. 
The recovered noise levels are given in Table~\ref{table:polaris}. 
They are comparable with the expected sensitivity for two repeats.

The dark blue dots in Fig.~\ref{fig:powspec250} show the power spectrum of the 250~$\mu$m map,
noise removed and divided by the transfer function $\phi(k)$ estimated using 
the official SPIRE beam profiles obtained on observations
of Neptune. We emphasize here that the SPIRE beam shapes cannot be approximated by a Gaussian for the level of precision
needed in this power spectrum analysis. Not taking into account the secondary lobes of the transfer 
function with a Gaussian would produce an artificial break in the power spectrum
at scales of 0.1-0.2~arcmin$^{-1}$ with a steepening of the slope at small scales.

Once corrected for noise and $\phi(k)$ the power spectrum shows a rather straight power law with
a slight flattening at wavenumbers larger than 1~arcmin$^{-1}$, 
typical of a white component due to point sources and the unresolved 
cosmic infrared background (CIB) Poissonian fluctuations.
{To extract the interstellar contribution to the power spectrum
we fitted the dark blue curve with $P_{sky}(k) = A_{ISM}k^{\gamma} + P_0$, 
where $P_0$ is the white level due to point sources and the CIB. 
The fit was done on scales between k=0.025~arcmin$^{-1}$, to exclude the largest scales 
where the IRIS-SPIRE comparison showed significant differences, and about twice the 
FWHM (2.0, 1.2 and 0.86 arcmin$^{-1}$ at 250, 350 and 500~$\mu$m respectively) to exclude
scales contaminated by residual noise.}
The recovered $\gamma$ and $P_{0}$ are given in Table~\ref{table:polaris}.
The uncertainties quoted in Table~\ref{table:polaris} 
include both the statistical error of the fit and a bias of a $\pm$10\% uncertainty on $N$. 
The increase of the uncertainty for $\gamma$ from 250 to 500~$\mu$m is due to the range of 
scales on which the fit is done, which get shorter with increasing beam size. 
Even though the uncertainties for $P_{0}$ are significant, it is remarkable that the levels obtained 
agree very well with the predictions of \cite{fernandez-conde2008} for the CIB fluctuation level at the SPIRE wavelengths.

At 250~$\mu$m the slope of the power spectrum is $\gamma=-2.65\pm0.10$ at wavenumbers from $k$=0.025 to 2.0~arcmin$^{-1}$.
In the top panel of Fig.~\ref{fig:powspec250} the green dots represent the power spectrum (multiplied by $k^{2.65}$)
of the SPIRE 250~$\mu$m map corrected for $N$, $\phi$ and $P_0$, leaving only the interstellar contribution. 
Within the uncertainties, the slopes are similar at all wavelengths (see Table~\ref{table:polaris}).
They are also compatible with the IRAS/IRIS 100~$\mu$m power spectrum measured on $12^\circ \times 12^\circ$ region centered 
on the SPIRE field\footnote{The IRAS/IRIS power spectrum shown here has been corrected for noise, 
the transfer function and the CIB
fluctuation level. It was scaled to match the SPIRE power spectrum at 10-15~arcmin scales.} (red dots).
However we note that the SPIRE power spectrum is systematically lower than IRAS for the first few large scale modes, in
agreement with what was shown in Sect.~\ref{section:iras}.
{This effect, which could be partly due to suppresion of large scale modes in the SPIRE map-making, 
certainly deserves further investigation especially by comparing data in the common bands of SPIRE and Planck.}

\section{Discussion and conclusions}

The power spectrum analysis of the 250, 350 and 500~$\mu$m SPIRE maps of the Polaris flare described here
shows that the -2.7 slope measured at large scales using IRAS \cite[]{miville-deschenes2007a} 
or CO data \cite[]{stutzki1998} extends by more than one order of magnitude in scales, down to 30''. 
{The high-resolution data obtained here on a 10 square degree field are in accordance with the results of
\cite{heithausen1998}, who also found a slope of -2.7 on a similar range of scales
combining high-resolution IRAM data of two 6' by 8' fields 
with lower resolution CfA and KOSMA data of a larger field.}
The analysis also reveals a Poissonian component at a level that agrees with CIB estimates.

As Polaris is uniformely heated by the interstellar radiation field, the dust grain temperature is 
likely to be rather uniform over the field. In these conditions brightness flutctuations in 
the SPIRE bands are dominated by dust column density fluctuations and the power spectrum of 
dust emission probes directly the power spectrum of the dust volume density in three dimensions.
Assuming gas and dust are well mixed in this diffuse cloud, the power spectrum of the gas volume 
density would then have a slope of $-2.7\pm0.1$ down to 0.01~pc. 
{The constant slope of the power spectrum suggests that the turbulent cascade is still
the main agent organizing the structure of the ISM at the 0.01~pc scale.}

Finally it is important
to point out that at this early phase of the {\it Herschel} mission it is impossible to rule out completely
that the analysis is affected at some level by instrumental or processing effects. 
{Nevertheless the comparison with IRAS/IRIS~100$\mu$m and the use of the psf measured in-flight provide
a coherent picture of the power spectrum analysis and show the capabilities of SPIRE at
performing mapping of diffuse emission.}

\begin{acknowledgements}
SPIRE has been developed by a consortium of institutes led by
Cardiff University (UK) and including Univ. Lethbridge (Canada);
NAOC (China); CEA, OAMP (France); IFSI, Univ. Padua (Italy); 
IAC (Spain); Stockholm Observatory (Sweden); Imperial College London,
RAL, UCL-MSSL, UKATC, Univ. Sussex (UK); and Caltech/JPL, IPAC,
Univ. Colorado (USA). This development has been supported by
national funding agencies: CSA (Canada); NAOC (China); CEA,
CNES, CNRS (France); ASI (Italy); MCINN (Spain); Stockholm
Observatory (Sweden); STFC (UK); and NASA (USA).
\end{acknowledgements}

\bibliographystyle{natbib}

 \begin{figure*}
 \begin{center}
 \includegraphics[width=12cm, draft=false, angle=0]{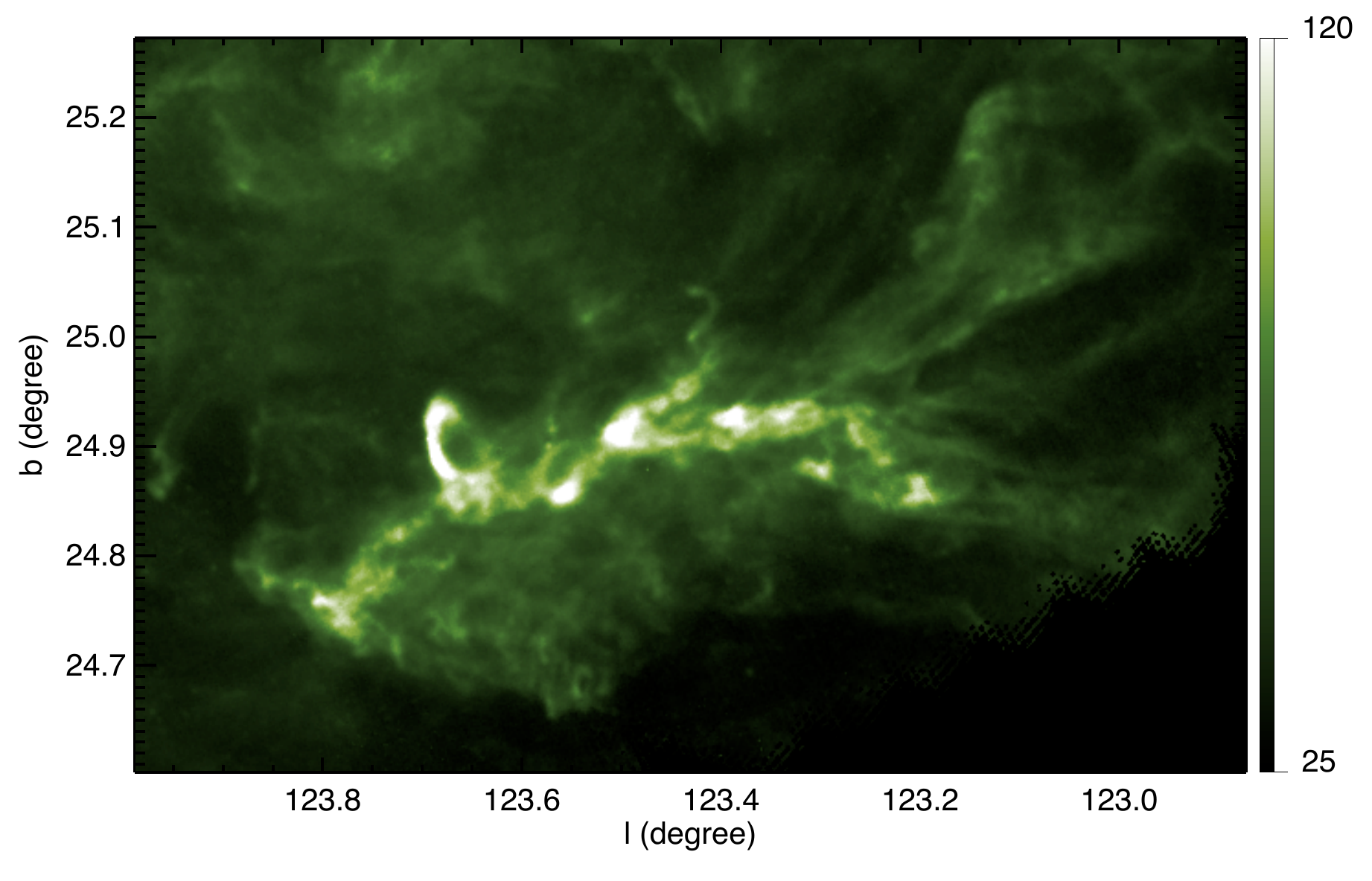}
 \includegraphics[width=12cm, draft=false, angle=0]{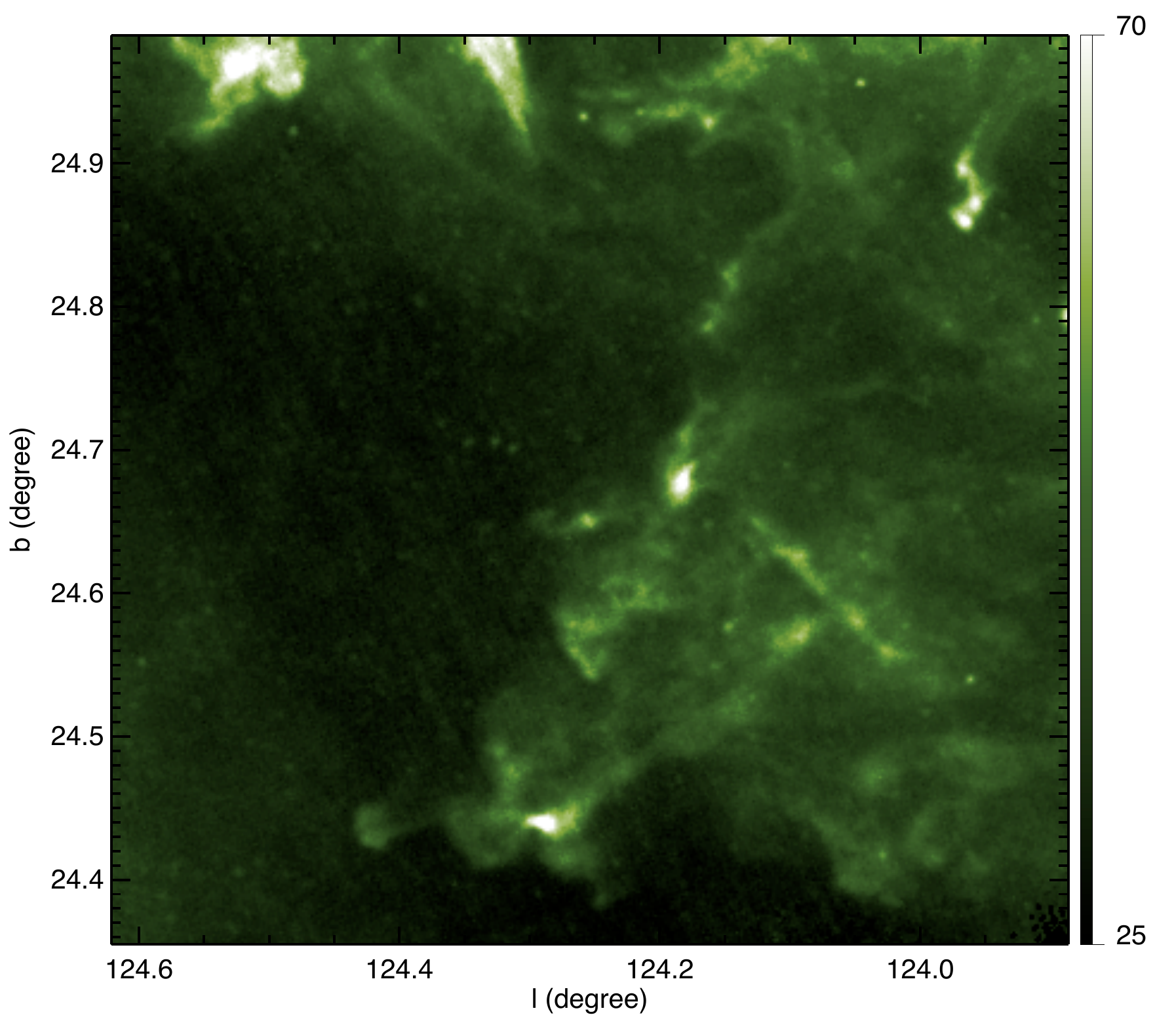}
 \caption{Zoom on specific regions of the Polaris flare. Spire 250 $\mu$m data in MJy sr$^{-1}$. }
 \end{center}
 \end{figure*}

 \begin{figure*}
 \begin{center}
 \includegraphics[width=12cm, draft=false, angle=0]{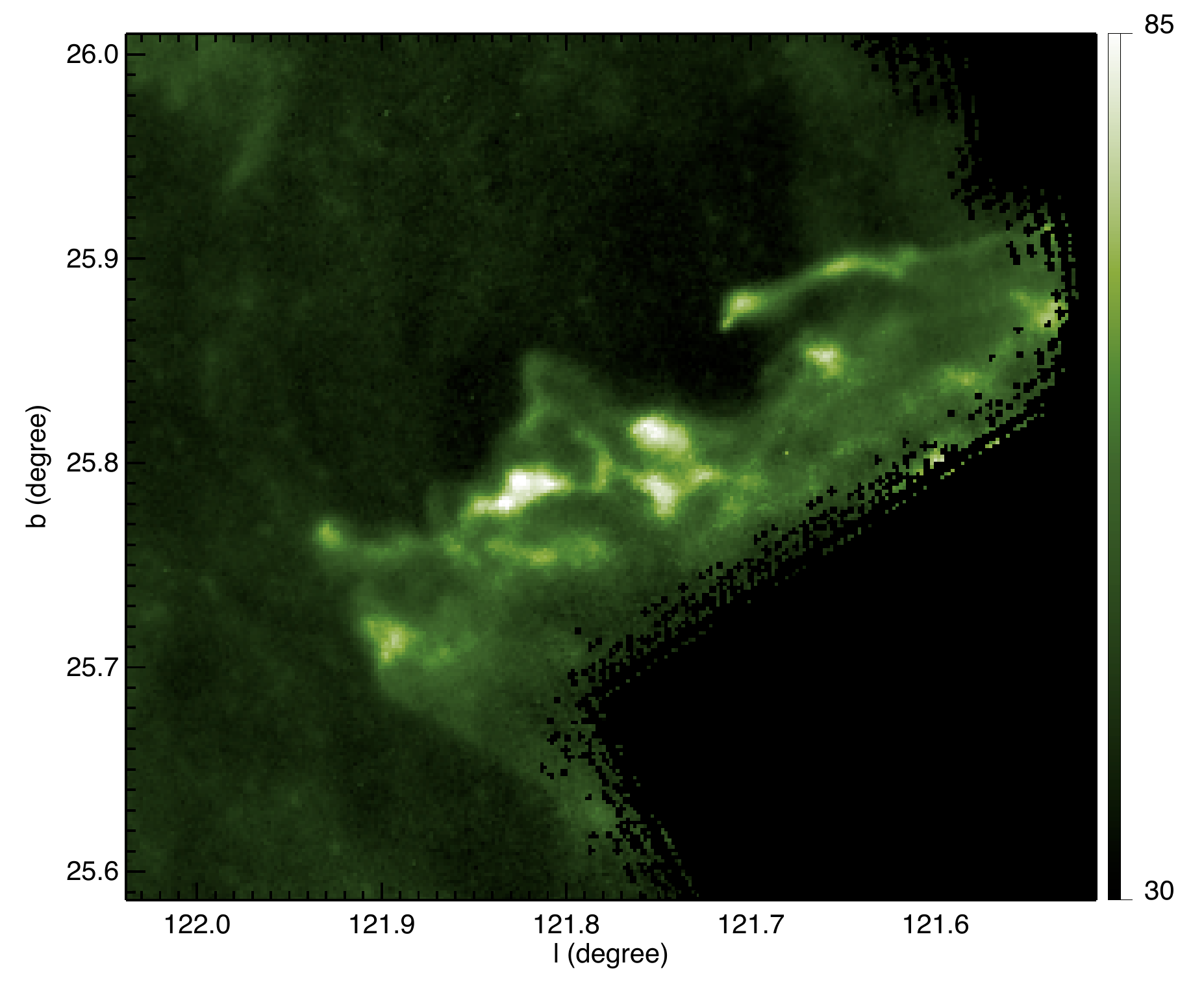}
 \includegraphics[width=12cm, draft=false, angle=0]{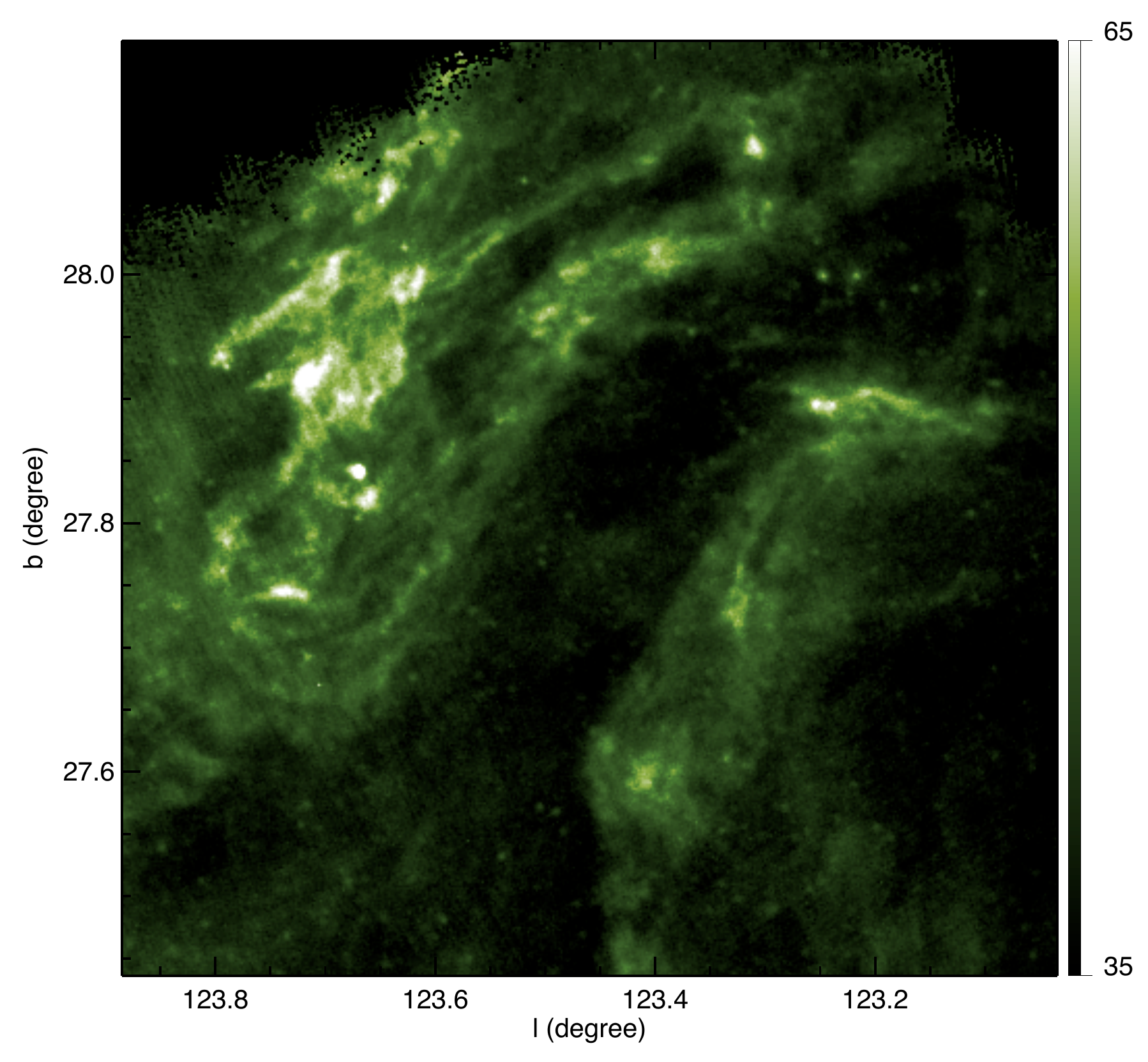}
 \end{center}
 \caption{Zoom on specific regions of the Polaris flare. Spire 250 $\mu$m data in MJy sr$^{-1}$.}
 \end{figure*}

 \begin{figure*}
 \begin{center}
 \includegraphics[width=12cm, draft=false, angle=0]{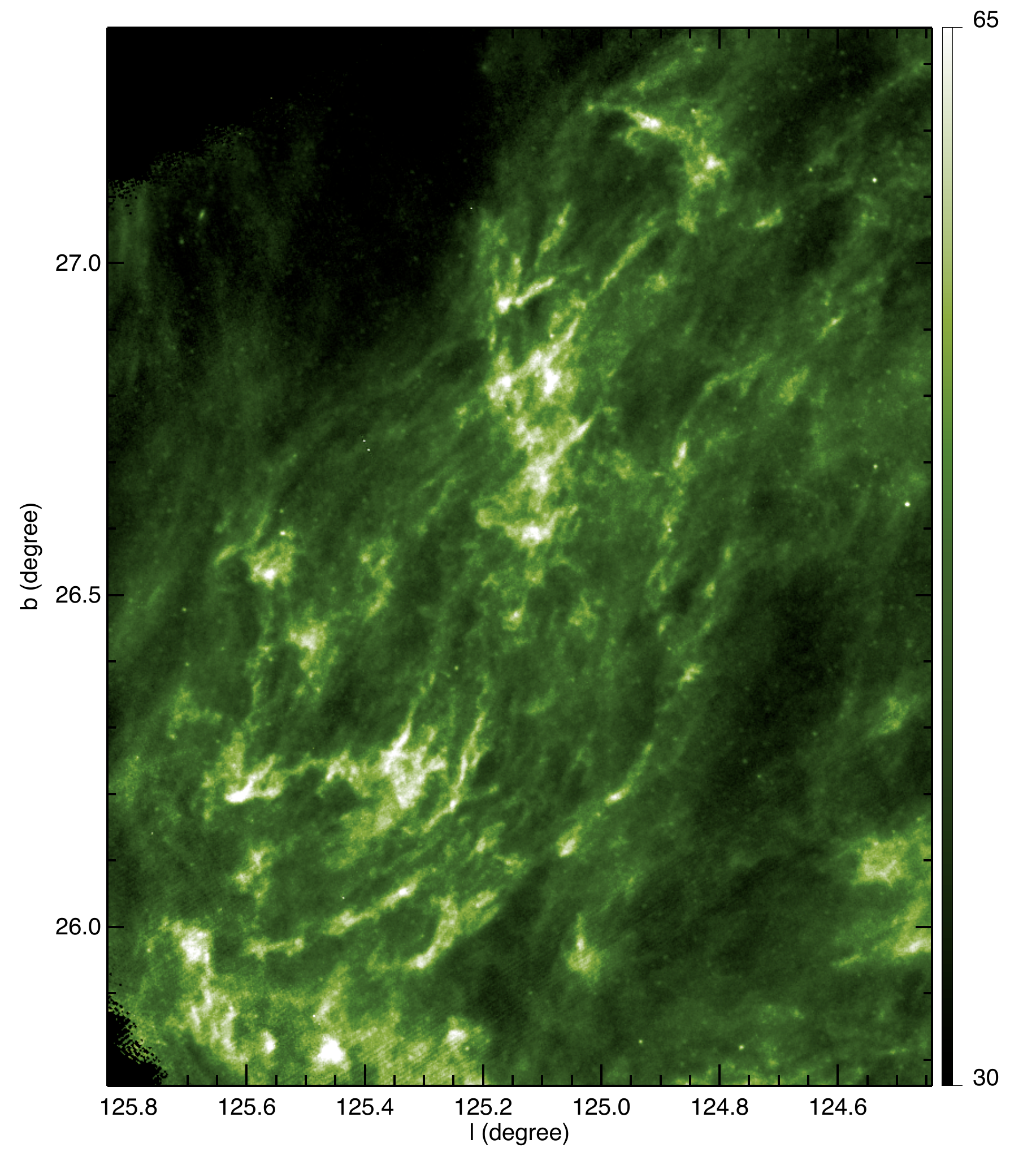}
 \end{center}
 \caption{Zoom on specific regions of the Polaris flare. Spire 250 $\mu$m data in MJy sr$^{-1}$.}
 \end{figure*}

\end{document}